# Magnetic Core-Shell Structure and Proximity Effect in 7 nm Single-Crystal $Co_3O_4$ Nanowires

Ping Lv[1], Yan Zhang[2], Rui Xu[1], Jia-Cai Nie[1], Lin He[1]*

[1] Department of Physics, Beijing Normal University, Beijing, 100875, People's Republic of China
[2] School of Physics, Peking University, Beijing, 100871, People's Republic of China.

We present a study of magnetic properties of single-crystal $Co_3O_4$ nanowires with diameter about 7 nm. The nanowires expose (111) planes composed of plenty of $Co^{3+}$ cations and exhibit two Néel temperatures at 56 K ($T_N$ of wire cores) and 73 K ($T_N$ of wire shells), which are far above $T_N$ = 40 K of bulk $Co_3O_4$. This novel bahavior is attributed to symmetry breaking of surface $Co^{3+}$ cations and magnetic proximity effect. The nanowire shells show macroscopic residual magnetic moments. Cooling in a magnetic field, a fraction of the residual moments are tightly pinned to the antiferromagnetic lattice, which results in an obvious horizontal and vertical shift of hysteresis loop. Our experiment demonstrates that the exchange bias field $H_E$ and the pinned magnetic moments $M_{pin}$ follow a simple expression $H_E = aM_{pin}$ with $a$ a constant.

The studies of the exchange bias was starting with the pioneering work of Meiklejohn and Bean in 1956 when studying Co particles embedded in their native antiferromagnetic oxide [1,2]. Since that time, there has been a continuously growing interest in both the fundamental physics and applications of this effect [3-5]. Several decades later, new phenomena, such as positive exchange bias and exchange bias in nanoscale antiferromagnetic (AFM) system, are still being uncovered in this subject [6-17].

In this letter, we present a study of magnetic properties of single-crystal AFM $Co_3O_4$ nanowires with diameter about 7 nm. The nanowires are selectively exposing four (111) planes composed of plenty of $Co^{3+}$ cations according to scanning tunnelling microscopy (STM) studies [18-20]. Remarkably, the nanowires show two characteristic order temperatures at 56 K and 73 K, which are far above the AFM order temperature of bulk $Co_3O_4$ ~ 40 K. The temperature 73 K is attributed to the Néel temperature of the nanowire shell with symmetry breaking of surface $Co^{3+}$ cations. The enhancement of $T_N$ of AFM core from about 40 K to 56 K is tentatively attributed to core-shell exchange interactions, *i.e.*, the so-called magnetic proximity effect proposed in Ref. [15,17]. The $Co_3O_4$ nanowire shell shows macroscopic residual magnetic moments below 35 K. Cooling the nanowires in a magnetic field larger than 20 kOe, up to 16% of the residual moments is tightly pinned to the antiferromagnetic lattice and does not rotate in an external magnetic field, which results in an obvious horizontal (exchange bias field $H_E$) and vertical shift of hysteresis loop. We show that the exchange bias field $H_E$ and the size of the pinned magnetic moments $M_{pin}$ follow a simple expression $H_E = aM_{pin}$ with $a$ a constant.

The bulk $Co_3O_4$ has a cubic spinel structure with lattice constant of 0.8065 nm. It is a semiconductor with the band-gap of 1.5 eV [19]. The $Co^{2+}$ cations of $Co_3O_4$ locate at the tetrahedral sites and the $Co^{3+}$ cations locate at the octahedral sites [21]. The magnetic moments of the $Co^{2+}$ cations (~4.14 $\mu_B$ per $Co^{2+}$ cation) exhibit AFM ordering at temperature $T < T_N$ = 40 K. The crystal field of the octahedral symmetry splits up the 3d orbitals into the $e_g$ and $t_{2g}$ levels. Due to large energy gap between $e_g$ and $t_{2g}$ (~ 2.0 eV) induced by octahedral symmetry, the six d-electrons of $Co^{3+}$ ($3d^6$) cation fill up the three lower energy $t_{2g}$ levels and therefore the $Co^{3+}$ cation of $Co_3O_4$ has zero magnetic moment. Recently, the magnetic properties of nanoscale $Co_3O_4$ have attracted much attention [22-26]. With decreasing the size of $Co_3O_4$, the ratio of surface-to-bulk increases and more $Co^{3+}$ cations with breaking of octahedral symmetry locate at the surface. As a result, it is expected that nanoscale $Co_3O_4$ shows novel magnetic properties, which are distinct from that of bulk phase. In previous papers, we synthesized ultrafine $Co_3O_4$ nanowires by heating cobalt foils under atomspheric conditions [18-20]. In this letter, we study the magnetic properties of the $Co_3O_4$ nanowires with average diameter about 7 nm. Detailed characterizations of the sample are reported in Ref. [18-20]. Fig. 1 shows a typical high-resolution transmission electron microscopy (HRTEM) image of the nanowires. The nanowires grow along the [110] direction and expose four (111) planes composed of plenty of $Co^{3+}$ cations [18-20].

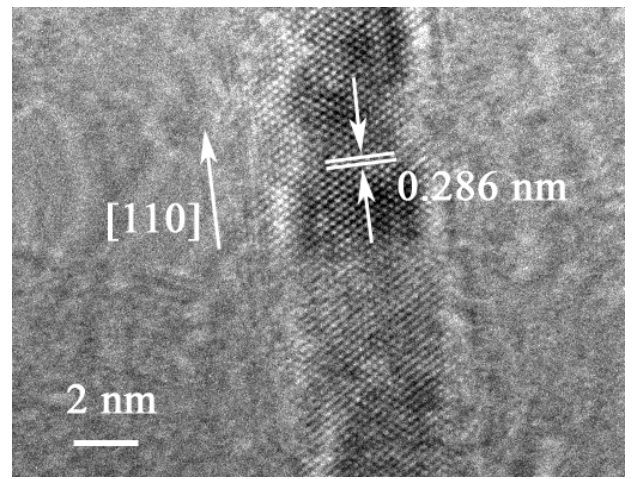

**FIG. 1**. A typical HRTEM image of the nanowire. The average diameter of the nanowires is 7 nm and the length 1-2 μm.

The magnetic properties were studied using a Quantum Design superconducting quantum interference device (SQUID). Fig. 2 shows temperature dependence of magnetization, $M(T)$ curves, under zero-field-cooling (ZFC) and field-cooling (FC) modes. The bifurcation of the ZFC and FC curves at low temperature indicates that there is an irreversible contribution, which will be elaborated later. Two clear peaks are observed at $T_1 \sim 73$ K and $T_2 \sim 56$ K in the ZFC curve and only one weak peak at about 70 K in the FC curve, as shown in inset (I) of Fig. 2(a). The positions of the two peaks $T_1$ and $T_2$ in the ZFC curve appear to be independent of the magnetic fields, as shown in Fig. 2(b). The lacking of the magnetic field dependence excludes both blocking temperature of fine ferromagnetic particle and freezing temperature of spin-glass as the origin of the peaks. These characteristic temperatures decrease with increasing magnetic field [27-29]. The observed features in our experiment are suggestive of the AFM phase transitions [23]. Similar two-peak structure was also observed by Benitez, *et al.* in $Co_3O_4$ nanostructures with size smaller than 8 nm and was attributed to the core-shell behavior of $Co_3O_4$ nanostructures [26]. However, in their experiments, the observed peaks appear at the temperatures that are much lower than the AFM order temperature of bulk $Co_3O_4 \sim 40$ K. Obviously, these peaks are distinct from that observed in our system.

We attribute the $T_1 \sim 73$ K to the AFM phase transition of the nanowire shell. The enhanced Néel temperature arises from the symmetry breaking of surface $Co^{3+}$ cations. In our experiment, the nanowires expose four (111) planes composed of $Co^{3+}$ cations [18-20]. The breaking of octahedral symmetry influences the crystal field and further results in "net" spin moment of surface $Co^{3+}$ cations. The appearance of magnetic moment of surface $Co^{3+}$ cations then changes the magnetic structure and correspondingly the magnetic properties of the $Co_3O_4$ nanowire shell. Our experimental result suggests that the appearance of magnetic moment of surface $Co^{3+}$ cations enhances the AFM coupling between $Co^{2+}$ cations and consequently increases the AFM order temperature of the nanowire shell. Due to the magnetic core-shell exchange interactions, the Néel temperature of the $Co_3O_4$ AFM core was also enhanced from about 40 K to $T_2 \sim 56$ K. Recently, similar magnetic proximity effect was proposed in Ref. [15,17] and observed in antiferromagnetic/ferrimagnetic core-shell nanoparticles.

The inset (II) of Fig. 2(a) shows inverse of susceptibility $1/\chi$ of the ZFC curve as a function of temperature. It exhibits a well-behaved linear property above 85 K. The dashed line is the fitting result to the data by the Curie-Weiss law, $\chi = C/(T-T_\theta)$, where $C = N\mu_0\mu_{eff}^2/3k_B$ with $N \sim 2.5 \times 10^{21}$ the number of formula per gram, $\mu_{eff}$ the effective magnetic moment per formula, $k_B$ the Boltzmann constant. $T_\theta$ is the Curie-Weiss temperature, which can be obtained from the extrapolation of the fitting line to $1/\chi = 0$. The sign of $T_\theta$ represents the type of exchange coupling, either AFM or ferromagnetic (FM), between magnetic moments. Usually, $T_\theta < 0$ for AFM coupling and $T_\theta > 0$ for FM coupling [30]. Interestingly, we obtained $T_\theta = 18$ K $> 0$. The slightly positive $T_\theta$ may arise from the FM coupling between magnetic moment of surface $Co^{3+}$ and $Co^{2+}$ cations. This novel behavior further confirms that the $Co_3O_4$ nanowire becomes a novel antiferromagnet due to the symmetry breaking of surface $Co^{3+}$ cations.

To further explore the magnetic properties of the $Co_3O_4$ nanowires, we carried out $M(H)$ measurements at various temperatures. Fig. 3 shows serveral typical $M(H)$ curves measured at 5, 10, 20, and 30 K. The high field linear part of the $M(H)$ curves, which arises from AFM contribution, was subtracted. The saturated magnetization $M_S$ reaches about 0.20 emu/g at 5 K. These macroscopic magnetic moments are the origin of irreversibility observed in the $M(T)$ curves under ZFC and FC models, as shown in Fig. 2(a). There are two main contributions of these macroscopic residual moments. One is the magnetic moments of the exposed $Co^{3+}$ cations at the surface, the other is the uncompensated spins of the $Co^{2+}$ cations. According to Néel' model [31,32], small AFM systems possess net magnetic moments from uncompensated spins and the number of uncompensated spins per particle can be estimated by $N_{UC} = (N)^q$ with $N$ the total number of magnetic atoms per particle, $q = 1/3$, $1/2$, and $2/3$ for certain crystalline orientations and particle morphologies. The magnetic structure of $Co^{2+}$ cations in our $Co_3O_4$ nanowires is more close to the case $q = 1/3$ in Néel' model. Taken the diameter of the nanowire ~ 7 nm and the length ~ 1 μm, the average volume of one nanowire is estimated as $4.9 \times 10^4$ nm$^3$, which implies the number of $Co^{2+}$ cations per particle $N \sim 7.5 \times 10^5$. The magnetization arising from the uncompensated spins ($N^{1/3} \sim 90$ $Co^{2+}$ cations per nanowire) is calculated as ~ 0.02 emu/g. This is only 10% of the experimentally determined value, which suggests that 90% of the magnetic moments arise from the surface $Co^{3+}$ cations (~ $2.9 \times 10^5$ surface $Co^{3+}$ cations per nanowire). The saturated magnetization decreases quickly with increasing temperature, as shown in the inset of Fig. (3), possibly due to the two-dimensional nature of the macroscopic residual moments.

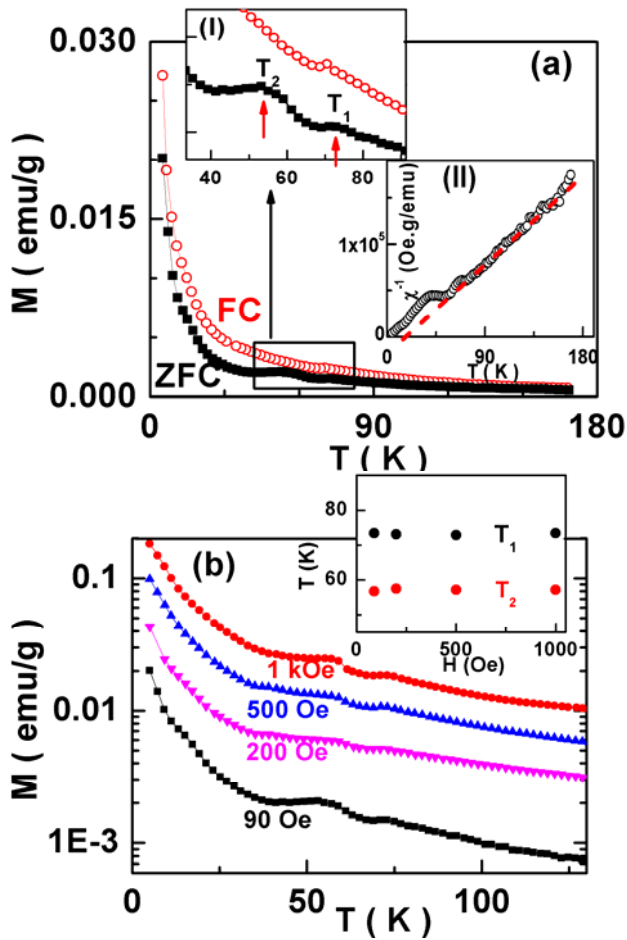

**FIG. 2**. (a) $M$ vs $T$ curves under ZFC and FC models measured in 90 Oe. The inset (I) is the enlarged view of the peaks, $T_1 \sim 73$ K and $T_2 \sim 56$ K, in the ZFC and FC curves. The inset (II) shows inverse of susceptibility $1/\chi$ as a function of temperature. The dashed line is the fitting result to the data by the Curie-Weiss law. (b) ZFC $M(T)$ curves measured in various applied fields from 90 Oe to 1000 Oe. The vertical coordinate is presented in the log-Y scale. The inset shows the characteristic temperatures, $T_1$ and $T_2$ in the ZFC curves, as a function of magnetic field.

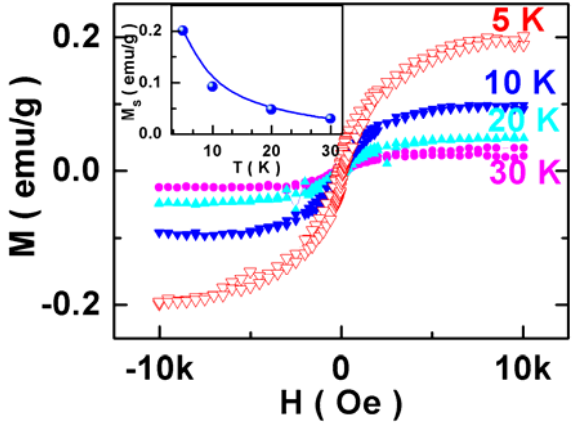

**FIG. 3**. $M(H)$ curves from -10 kOe to 10 kOe measured at various temperatures. The high field linear part of the $M(H)$ curves was subtracted. The inset shows the saturated magnetization $M_S$ as a function of temperature.

In order to further identify the magnetic core-shell exchange interactions of the nanowires, *i.e.*, the magnetic proximity effect, more magnetometry studies were performed. Fig. 4(a) shows ZFC and FC hysteresis loops measured at 5 K. The high field linear part of the $M(H)$ curves was subtracted. The original recorded $M(H)$ curves are shown in inset (I). For the FC hysteresis loop, the sample is cooled in a cooling field $H_{FC} = 20$ kOe from 100 K ($> T_1$ and $T_2$) to 5 K. The FC and ZFC $M(H)$ curves are superpositioned in high positive field and show an obvious bifurcation $\Delta M = 0.062$ emu/g in high negative field. The vertical shift $\Delta M$ is due to pinned magnetic moments $M_{pin}$ that are not rotated by the applied field. In the FC process, a fraction of magnetic moments $M_{pin}$ are tightly pinned to the antiferromagnetic lattice along the direction of the cooling field [2-5,10]. As a result, the saturated magnetization is $M_S$ in positive magnetic field and ($M_S$ - $2M_{pin}$) in negative field, which results in the vertical shift $\Delta M = 2M_{pin}$ of the hysteresis loop. It indicates that we can obtain $M_{pin}$ and the ratio of the pinned magnetic moments **Pin** = $M_{pin}/M_S \sim 16\%$ (with $H_{FC} = 20$ kOe) of the nanowires by magnetic measurements. The core-shell interfacial exchange interactions also lead to the exchange bias field $H_E$ and the enhanced coercivity $H_C$ of the nanwires, as shown in inset (II) of Fig. 4 (a). In the ZFC $M(H)$ curve, the coercivity $H_C$ is only 130 Oe. Whereas, in the FC $M(H)$ curve with $H_{FC} = 20$ kOe, the coercivity $H_C$ is enhanced to 270 Oe and we observe a exchange bias field $H_E = 375$ Oe.

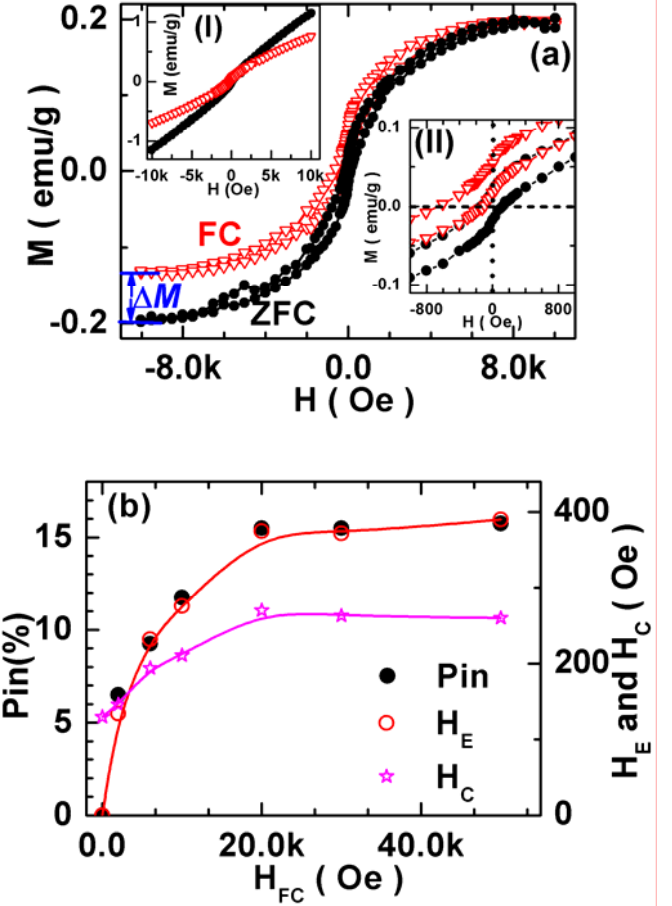

**FIG. 4**. (a) $M(H)$ curves at 5 K under ZFC and FC modes. For the FC hysteresis loop, the sample is cooled in $H_{FC} = 20$ kOe from 100 K to 5 K. The high field linear part of the $M(H)$ curves was subtracted. $\Delta M = 2M_{pin}$ represents the vertical shift of the hysteresis loop due to the pinning of the residual magnetic moments. The inset (I) shows the original recorded $M(H)$ curves under ZFC and FC modes. The pinned moments to the antiferromagnetic lattice also influence the magnetic properties of antiferromagnet, *i.e.*, the slope of high field linear part is changed in ZFC and FC curves. The inset (II) shows the $M(H)$ curves in the low field region. Clear exchange bias can be observed in the FC $M(H)$ curve. (b) The left Y-axis shows the ratio of pinned magnetic moments **Pin** = $M_{pin}/M_S$ at 5 K as a funcion of the cooling field $H_{FC}$. The right Y-axis shows the exchange bias field $H_E$ and coercivity $H_C$ at 5 K as a function of cooling field $H_{FC}$. The solid curves are guide to eyes. We obtain a simple and quantitative relation $H_E = aM_{pin}$ with $a$ a constant

Fig. 4(b) shows the ratio of pinned magnetic moments **Pin** (left Y-axis), the exchange bias field $H_E$ (right Y-axis), and the coercivity $H_C$ (right Y-axis) as a funcion of the cooling field $H_{FC}$ measured at 5 K. The coercivity $H_C$ increases steady from 130 Oe with increasing the cooling field until reaching a constant ~ 270 Oe when the cooling field is larger than 20 kOe. Similarly, the ratio of pinned magnetic moments **Pin** and the exchange bias field $H_E$ also increase with increasing the cooling field and become saturated above $H_{FC}$ ~ 20 kOe. It is well accepted that the exchange bias arises from a fraction of pinned FM moments into a well-defined reference direction by an antiferromagnet [3-5]. Whereas, a quantitative link between them is still lacking. A remarkable result obtained in our experiment is that the ratio of pinned magnetic moments **Pin** and the $H_E$ show almost the same dependencies on the cooling field, *i.e.*, **Pin** = $M_{pin}(H_{FC})/M_S \propto H_E(H_{FC})$. Therefore, we conclude that the exchange bias field $H_E$ and the size of the pinned magnetic moments $M_{pin}$ follow a simple and quantitative expression $H_E(H_{FC}) = aM_{pin}(H_{FC})$ with $a$ a constant.

In literature, many models were proposed to explain the size of exchange bias [3-5]. However, the quantitative link between theory and experiments remain forefront research problems due to the lacking of detailed interfacial information. Here we semiquantitative estimate the exchange bias field of the $Co_3O_4$ nanowire following the random field model [33]

$$H_E = \frac{2z\sqrt{AK}}{\pi^2 M_F t_F}. \qquad (1)$$

Here, z is the number of nearest neighbouring magnetic moments, $A$ and $K$ the exchange stiffness and anisotropy energy of the antiferromagnet respectively, $t_F$ the thickness of the ferromagnetic layer. The exchange bias field is estimated as about 1 kOe [34], which is not far from our experimental result.

In summary, we present a study of magnetic core-shell structure and magnetic proximity effect in 7 nm single-crystal $Co_3O_4$ nanowires. Our experiment indicates that the novel magnetic properties of the nanowires arise from symmetry breaking of surface $Co^{3+}$ cations. The $Co_3O_4$ nanowire shell shows macroscopic residual magnetic moments mainly from the contribution of surface $Co^{3+}$ cations. Cooling the nanowires in a magnetic field, a fraction of the residual moments is tightly pinned to the antiferromagnetic lattice, which results in an obvious horizontal and vertical shift of hysteresis loop. Remarkably, our experiments demonstrate that the exchange bias field $H_E$ and the size of the pinned magnetic moments $M_{pin}$ follow a simple and quantitative expression $H_E = aM_{pin}$.

Ping Lv and Yan Zhang contributed equally to this paper. This work is supported by National Natural Science Foundation of China (Nos. 11004010 and 10974019), the Fundamental Research Funds for the Central Universities.

*Correspondence to: helin@bnu.edu.cn